\documentclass[twocolumn]{aastex631}

\usepackage{graphicx}
\usepackage[none]{hyphenat}
\usepackage{amsmath}
\usepackage{booktabs}
\usepackage{multirow}
\usepackage{txfonts}
\usepackage{enumitem}
\usepackage{algorithmic}
\usepackage{bm}
\usepackage{mathrsfs}
\usepackage{booktabs}  
\usepackage{float}
\usepackage{subfigure}
\DeclareRobustCommand{\VAN}[3]{#2}
\let\VANthebibliography\thebibliography
\def\thebibliography{\DeclareRobustCommand{\VAN}[3]{##3}\VANthebibliography}
\usepackage{color}

\begin{document}

\title{Magnetic reconnection-driven turbulence and turbulent reconnection acceleration}

\author{Shi-Min Liang}
\affiliation{Department of Physics, Xiangtan University, Xiangtan, Hunan 411105, People’s Republic of China;\\}

\author{Jian-Fu Zhang}
\affiliation{Department of Physics, Xiangtan University, Xiangtan, Hunan 411105, People’s Republic of China;\\}
\affiliation{Key Laboratory of Stars and Interstellar Medium, Xiangtan University, Xiangtan 411105, People’s Republic of China\\}

\author{Na-Na Gao}
\affiliation{Department of Physics, Xiangtan University, Xiangtan, Hunan 411105, People’s Republic of China;\\}

\author{Hua-Ping Xiao}
\affiliation{Department of Physics, Xiangtan University, Xiangtan, Hunan 411105, People’s Republic of China;\\}
\affiliation{Key Laboratory of Stars and Interstellar Medium, Xiangtan University, Xiangtan 411105, People’s Republic of China\\}
\email{jfzhang@xtu.edu.cn (JFZ), hpxiao@xtu.edu.cn (HPX)}


\begin{abstract}
This paper employs an MHD-PIC method to perform numerical simulations of magnetic reconnection-driven turbulence and turbulent reconnection acceleration of particles. Focusing on the dynamics of the magnetic reconnection, the properties of self-driven turbulence, and the behavior of particle acceleration, we find that: (1) when reaching a statistically steady state of the self-driven turbulence, the magnetic energy is almost released by 50\%, while the kinetic energy of the fluid increases by no more than 15\%. (2) the properties of reconnection-driven turbulence are more complex than the traditional turbulence driven by an external force. (3) the strong magnetic field tends to enhance the turbulent reconnection efficiency to accelerate particles more efficiently, resulting in a hard spectral energy distribution. Our study provides a particular perspective on understanding turbulence properties and turbulent reconnection-accelerated particles.

\end{abstract}


\section{Introduction}  \label{section1}
Magnetic reconnection has been widely regarded as one of the important mechanisms for the efficient acceleration of non-thermal high-energy particles and is a hot research topic in space physics and astrophysics. Magnetic reconnection is a topological recombination of magnetic fields, accompanied by violent energy release. The phenomenon of reconnection widely exists in various astrophysical environments, such as solar flares (\citealt{Dere1996,Ciaravella2008,Chitta2020}), pulsar wind nebula (\citealt{Meyer2010, Tavani2011}) and gamma-ray bursts (\citealt{Gehrels2009}).

Magnetic reconnection occurs when magnetic field lines with opposite polarity encounter each other. The magnetic field lines annihilate at the discontinuity position to form a current sheet. In the case of limited resistivity, the reconnection rate is constrained by the resistivity \citep{Parker1957, Sweet1958}. In the Sweet-Parker model, the reconnection speed is given by $V_{\rm R} \sim V_{\rm A}(\Delta/L) \sim V_{\rm A} S^{-1/2} \ll 1$, where $L$ and $\Delta$ are the length and thickness of the reconnection layer, respectively. The Alfv\'en velocity is defined as $V_{\rm A}=B/\sqrt{4\pi \rho}$ with the plasma density $\rho$, and the Lundquist number as $S = LV_{\rm A}/\eta$ with the resistive diffusivity $\eta$. In a realistic astrophysical environment, due to a large-scale $L$ and a small resistive diffusivity $\eta$, we usually have an extremely large $S$, predicting a slow reconnection speed $V_{\rm R} \ll V_{\rm A}$. This slow reconnection rate cannot explain the fast flare phenomena by \cite{Dere1996} who estimated the reconnection rate as being about 0.1 times the Alfv\'en velocity.

To solve the problem of slow reconnection rate, the `X-point' reconnection configuration was proposed by \cite{Petschek1964} who assumed that the length-scale $L$ is small, that is, $L$ is similar to $\Delta$ by bending the magnetic field toward the reconnection point at a sharp angle. This modification can improve the reconnection rate but the configuration is unstable, rapidly collapsing to the Sweet-Parker configuration as confirmed in the MHD numerical simulations (\citealt{Biskamp1996}). There exist some other solutions where the reconnection rate can be improved by constructing a special configuration of magnetic field (see \citealt{Priest2007} for more details). The main problem of these models is that they dealt with special magnetic field structures that may not happen in a general astrophysical environment. 

Later, Lazarian \& Vishniac (\citeyear{Lazarian1999}, henceforth LV99) proposed a turbulent magnetic reconnection model based on MHD turbulence theory (\citealt{Goldreich1995}, henceforth GS95). In this model, the outflow width $\Delta$ is not determined by microscopic diffusive processes but by the random walk of magnetic field lines. LV99 model predicted that the reconnection rate changes with the level of turbulence. In a turbulent medium, the wandering of the magnetic field lines allows for many simultaneous events of reconnection which make it fast. 
At the same time, the turbulence makes the reconnection region thicker, i.e., much larger $\Delta$. Both cases could improve the reconnection rate by $V_{\rm R} =V_{\rm A} {\rm min} [(l/L)^{1/2}, (L/l)^{1/2}] (v_l/V_{\rm A})^2$, where $v_l$ is the eddy velocity at the scale $l$. 
As for strong turbulence with $v_l\sim V_{\rm A}$, the reconnection rate can reach the Alfv\'en velocity $V_{\rm A}$ on the system size. The relevant theoretical predictions in LV99 have been numerically confirmed by \cite{Kowal2009}. This model makes the GS95 MHD turbulence theory more self-consistent. 
Note that LV99 model has been successfully applied to many astrophysical environments, such as solar physics (\citealt{Lazarian2009a}), active galactic nuclei (\citealt{Kadowaki2015}), black hole X-ray binaries (\citealt{deGouveiadalPino2005, deGouveiaDalPino2010}), gamma-ray bursts (\citealt{Lazarian2003, Lazarian2019}; \citealt{Zhang2011}; \citealt{Xu2017}), and galaxy clusters (\citealt{Brunetti2007, Brunetti2016}).

The intense release of magnetic energy caused by a high reconnection rate is considered as one of the effective ways to accelerate charged particles. At the same time, the magnetic energy released by magnetic reconnection also drives turbulence and heat particles. 
The reconnection acceleration of particles occurs due to the flow convergence driven by reconnection, regardless of the compressibility of the medium (\citealt{Lazarian2020}; \citealt{Xu2023}). In analogy to the classical diffusive shock (first-order Fermi) acceleration, where particles are confined in the vicinity of a shock, de Gouveia dal Pino \& Lazarian (\citeyear{deGouveiadalPino2005}, henceforth GL05) first proposed that the first-order Fermi process operates within the 3D turbulent reconnection region. Particles confined in the converging magnetic fluxes of opposite polarity bounce back and forth between the reconnection-driven inflows (\citealt{Kowal2011}). On the other hand, fast reconnection acceleration requires a sufficiently high reconnection rate and therefore a sufficiently high turbulence level. The presence of turbulence not only regulates the inflow velocity but also introduces various inflow inclinations relative to the local turbulent magnetic field (\citealt{Xu2023}).

Numerical simulation is the most effective way to understand magnetic reconnection on kinetic plasma and/or MHD scales. Earlier studies on reconnection acceleration mostly focused on 2D reconnection structures. Through 2D numerical simulation, \cite{Drake2006} reported that the first-order Fermi acceleration happens in the 2D reconnection. 
In the restricted 2D configuration, particles are trapped within shrinking magnetic islands, i.e., plasmoids, which can arise from the plasma tearing instability when a long narrow current sheet is prescribed (see \citealt{Drake2010,Sironi2014}). 
The trapping of particles within the islands limits the acceleration efficiency in 2D reconnection. It is generally accepted that 2D magnetic islands are not tenable and susceptible to turbulence. 3D simulation should be more realistic and close to a real astrophysical process. In the 3D case, the dynamics of magnetic field lines are very different, and turbulence makes the magnetic field lines more random (\citealt{Eyink2011}). Note that 3D reconnection acceleration is more efficient than its 2D counterpart (\citealt{ZhangH2021}; \citealt{ZhangQ2021}). At the same time, the appearance of the first-order Fermi process has been claimed on the kinetic plasma scale using 3D PIC simulations \citep{Guo2014, Guo2015}. 
On the macroscopic MHD scale, the testing of the GL05 picture was first performed by \cite{Kowal2011} using 3D turbulent reconnection simulations \citep{Kowal2009} together with test particle methods. By using an externally driving turbulence, \cite{Kowal2011,Kowal2012} claimed that the particles trapped within the reconnection region are accelerated in the form of the first-order Fermi acceleration, and demonstrated that the reconnection acceleration is efficient in the presence of turbulence. 
Without involving external force to drive turbulence, \cite{Beresnyak2017} \& \cite{Kowal2017} studied the self-driven turbulent magnetic reconnection. They confirmed that stochastic reconnection can cause turbulence which would further promote the development of magnetic reconnection. In the framework of reconnection-driven turbulence, \cite{Zhang2023} used the test particle method integrating the motion of charged particles to confirm the effective acceleration of particles.

From the perspective of numerical methods, the advantage of PIC methods is that the influence of micro dynamic scale can be observed, but not extend to a large scale due to the limited particle gyroradius. 
As for MHD-test particle methods, it can study the reconnection acceleration at the macroscopic scale, which is helpful to understand the overall process on the macroscopic scale without considering the micro-plasma effects. 
In this work, we adopt the MHD-PIC module (\citealt{Bai2015,Mignone2018}) from astrophysical simulation code PLUTO (\citealt{Mignone2007}). 
This hybrid MHD-PIC module is advantageous to explore the cosmic ray (CR) dynamic effect on a scale larger than the ion's skin depth (\citealt{Mignone2020}) (see Section \ref{MHD-PIC} for more details).

Focusing on the coevolution of particles and fluids, the purpose of our work is to understand the reconnection-driven turbulence and reconnection acceleration in the compressible MHD framework. 
We want to explore the properties of reconnection-driven turbulence and particle acceleration on the transition scale between the large MHD scale and the small kinetic plasma one. The layout of this paper is as follows. 
Our simulation methods are presented in Section \ref{SimuMethod}. Section \ref{StatMethod} describes the statistical methods used in this paper. In Section \ref{results}, we provide the results of the numerical simulation. Finally, we give the discussion and summary in Sections \ref{Discussion} and \ref{Summary}, respectively. 

\section{ Simulation Method}   \label{SimuMethod}
\subsection{ MHD-PIC module} \label{MHD-PIC}
The MHD-PIC module embedded in PLUTO code can simulate the evolution of fluids by solving the set of MHD equations and simultaneously integrate the motion of CR particles using conventional PIC techniques. Particle back-reaction on the fluid is included in the form of momentum–energy feedback, and the CR-induced Hall term in Ohm’s law was introduced by \cite{Bai2015}. The structure of this module mainly included three parts: solution of the MHD equations, integral of the motion equation of particles, and stepping time scheme. In general, this module, dealing with microphysics such as the resistance and Hall effect in the MHD part, is appropriate to capture the dynamical evolution of a plasma consisting of a thermal fluid and a non-thermal component represented by relativistic charged particles. Since the gyroradius of CRs (typically representing energetic protons) greatly exceeds the ion skin depth of the plasma, this module extends the range of applicability to much larger spatial (and longer temporal) scales.

The MHD-PIC module adopts finite-volume numerical schemes to solve the set of MHD equations on a computational grid, while the injected computational particles represent clouds of physical particles that are close to each other in phase space (\citealt{Mignone2018}). The divergence-free condition is enforced using either constrained transport or divergence-cleaning approaches. In addition, particles are treated kinetically either using conventional PIC time-reversible integrators or simple Runge-Kutta schemes. We notice that this module has been successfully tested for Bell instability and shock acceleration, and the results are in good agreement with previous studies based on Athena code (\citealt{Bai2015}). In the parallel computing tests, the CPU utilization efficiency can be stable at over 0.8 (\citealt{Mignone2018} for more details). 

\subsection{Simulation Setup}\label{SimuSetup}
To study turbulent magnetic reconnection processes, we perform numerical simulations employing the MHD-PIC module \citep{Mignone2018} mentioned above to solve the dimensionless ideal MHD equations as follows
\begin{equation} \label{1}
\frac{\partial \rho}{\partial t}+\nabla \cdot{(\rho \bm{v}_{\rm g})}=0,
\end{equation}
\begin{equation} \label{2}
\frac{\partial \bm{m}}{\partial t}+\nabla \cdot[\bm{m} \bm{v}_{\rm g} -\bm {BB}+I(p+\frac{\bm{B}^2}{2})]=0,
\end{equation}
\begin{equation} \label{3}
\frac{\partial E_{\rm t}}{\partial t}+\nabla \cdot [(\frac{1}{2}\rho \bm{v}_{\rm g}^2+\rho e+p)\bm{v}_{\rm g} + c\bm{E} \times \bm{B}]=0,
\end{equation}
\begin{equation} \label{4}
\frac{\partial \bm{B}}{\partial t}+\nabla \times(c\bm{E})=0,
\end{equation}
\begin{equation} \label{5}
\nabla \cdot \bm{B}=0.
\end{equation}
These equations are the continuity, momentum, energy, induction, and solenoidal condition, respectively. Here, $\rho$ is the mass density, $\bm{m}=\rho \bm{v}_{\rm g}$ the momentum density, $I$ unit tensor, $e$ specific internal energy, $\bm{v}_{\rm g}$ the gas velocity, $p$ the gas pressure, $\bm{B}$ magnetic field, and $E_{\rm t} = \rho e + \bm{m}^2/2 \rho+ \bm{B}^2/2$ the total energy density. The evolution of the magnetic field is governed by Faraday’s law of $\bm{E} =-\bm{v}_{\rm g} \times \bm{B}$. It should be stressed that we do not include resistant dissipation in our simulation.

Our numerical simulations are performed in a 3D domain with physical dimensions $1.0 L \times 0.5 L \times 1.0 L$, where the dimensionless length scale is set to $L = C/\omega_{\rm p}=10^4$ with the light speed of $C=10^4V_{\rm A}$ and the plasma frequency of $\omega_{\rm p} = \sqrt{4\pi \rho q^2/m_{\rm p}}$ ($q$, $m_{\rm p}$ are a charge and mass of the proton, respectively). The configuration of the initial magnetic field is considered as a Harris type by (\citealt{Harris1962})
\begin{equation} \label{8}
\bm{B} = B_0 {\rm tanh}\frac{2 \pi y}{w}\bm{e}_{\rm x} ,
\end{equation}
where $w$ is the initial width of the current sheet, and $B_0$ is the magnetic field strength controlled by parameter $\sigma = B_0^2 / (4 \pi \rho)$. The magnetic fields are antiparallel to the $X$-direction, resulting in a discontinuity placed on the $X-Z$ plane. By counteracting the Lorenz-force term with a thermal pressure gradient, we can obtain an initial equilibrium condition 
\begin{equation} \label{9}
p = \frac{(\beta +1)}{8\pi}B_0^2-\frac{B^2}{8\pi},
\end{equation}
where the plasma parameter is set as $\beta = 0.01$ \citep{Puzzoni2021}. As a result, the total pressure remains invariable throughout the whole current sheet.

The evolution of CR particles can be described by the following equations
\begin{equation}    \label {particle-equation1} 
    \frac{{\rm d}\bm{x}_{\rm p}}{{\rm d}t} = \bm{v}_{\rm p},
\end{equation}
\begin{equation}  \label{particle-equation2}
    \frac{{\rm d}(\gamma \bm v)_{\rm p}}{{\rm d} t} = \alpha_{\rm p}(c\bm E + {\bm v}_{\rm p}\times \bm B),
\end{equation}
where $\bm x_{\rm p}$ and $\bm v_{\rm p}$ indicate the spatial position and velocity of a charged particle (proton for our scenario), respectively. The parameter $\alpha_{\rm p}$ in Equation (\ref{particle-equation2}) denotes the particle charge to mass ratio.
When numerically solving Equations (\ref{particle-equation1}) and (\ref{particle-equation2}) by the Boris integrator (\citealt{Boris1971}), the electric and magnetic fields $\bm E$ and $\bm B$ are computed from the magnetized fluid by a cubic spline interpolation approach.

We uniformly inject $10^6$ particles into the whole box space and initialize them with a Maxwellian distribution of the thermal velocity of $0.1V_{\rm A}$ in each direction. The test particles are evolved together with the fluid using the Boris algorithm (\citealt{Mignone2018}). We use periodic boundaries in the $X$ and $Z$ directions and reflective conditions in the $Y$ direction. To numerically solve Equations (\ref{1}) to (\ref{5}), we choose the HLL Riemann solver with the Characteristic Tracing Contact (CT-Contact) electromagnetic fields averaging scheme and the second-order piecewise linear reconstruction and 2rd-order Runge Kutta time stepping. 
Except for the part on comparative studies of the resolution, we set the numerical resolution to $512 ~\times ~256 ~\times ~512$ through this paper, leading to the grid size of $\delta L \simeq 20$. 
In our simulations, we set some fixed parameters such as the uniform background plasma density of $\rho=1.0$, light speed of $C=10^4$, and Alfv\'en velocity of $V_{\rm A}=1.0$. 
The variable parameters are listed in Table \ref{table} where we set the initial velocity perturbation with the amplitude of $V_{\rm eps} \sim 0.1 V_{\rm A}$ to promote the generation of turbulence, that is, to shorten the simulation time. When the spectral energy distributions of the test particles reach a statistically steady state, we terminate the simulation at the final integration time of $t = 7\times 10^4 \omega_{\rm p}^{-1}$. 

\section{Statistical method} \label{StatMethod}
We adopt basic statistical tools such as power spectrum and structure-function to capture the statistical characteristics of reconnection-driven turbulence. The power spectrum can provide information concerning energy cascade processes in MHD turbulence and is described as
\begin{equation} \label{equation power spectrum}
    P_{n\rm {D}}(\bm{k}) = \frac{1}{(2\pi)^{n}}\int C(\bm{r})e^{-i\bm{k}\cdot \bm{r}}d\bm{r},
\end{equation}
where $n = 1,2,3$ denote the number of the dimension of physical space. The $C(\bm{r})$ of the integrand is a correlation function defined as 
\begin{equation} \label{equation correlation}
C(\bm{r}) = \langle A(\bm{r}_1)A(\bm{r}_2)\rangle,
\end{equation}
which is suitable for any physical quantity $A$. As usual, brackets denote a spatial average. Moreover, we obtain the shell-integrated 1D power spectrum for a 3D variable using a spherical shell integral (e.g., \citealt{zhang2018})
\begin{equation} \label{12}
    E_{\rm{3D}}(k) = \int_{k-0.5}^{k+0.5} P_{\rm{3D}}(k)dk,
\end{equation}
over a shell with inner $(k - 0.5)$ and outer $(k + 0.5)$ radii in 3D Fourier space.

The second-order structure function is defined as
\begin{equation} \label{equation SF}
    SF_2(R, z) = \langle |\bm{A}(r_1)-\bm{A}(r_2)|^2 \rangle,
\end{equation}
where ${R} = |\hat{z} \times (\bm{r}_2 - \bm{r}_1)|$, $z = \hat{z} \cdot (\bm{r}_2 - \bm{r}_1)$, $\hat{z} = \bm{A}_l/|\bm{A}_l|$. Here, $R$ and $z$ are coordinate components in a cylindrical coordinate system in which the $z$-axis is parallel to $\bm{A}_l$. Based on \cite{Cho2000}, we can calculate the local physical quantity at $\bm r$ by
\begin{equation} \label{equation local frame}
    \bm{A}_l(\bm{r}) = [\bm{A}(\bm{r}_1)+\bm{A}(\bm{r}_2)]/2,
\end{equation}
where $\bm{A}(\bm{r}_1)$ and $\bm{A}(\bm{r}_2)$ represents the local fields at any two 3D space locations. 

As for the test particles, we can analyze their gyroradii, spectral energy distributions, and momentum diffusion coefficients. The momentum diffusion coefficient is defined as 
\begin{equation} \label{equation dpp}
    D_{\rm{pp}} = \frac{\langle(\delta p)^2 \rangle}{2\delta t},
\end{equation}
where $\delta p$ represents the momentum changes at the time interval $\delta t$. 

\begin{table}
  \begin{center}
\setlength{\tabcolsep}{5mm}
    \begin{tabular}{ccccc} 
    \hline
      \text{Models} & \text{$\sigma$} & \text{$V_{\rm eps}$} & \text{$w$} & \text{$B_{\rm g}$}\\
    \hline
M1   & 4.0    & 0.1   & 200.0   & 0.0\\
M2   & 1.0    & 0.01  & 200.0   & 0.0\\    
M3   & 1.0    & 0.1   & 600.0   & 0.0\\
M4   & 1.0    & 0.1   & 200.0   & 0.2\\  
      \hline
    \end{tabular}
    \caption{Variable parameters used in our simulations. $V_{\rm eps}$ is the amplitude of initial velocity perturbation, $\sigma$ the magnetic parameter, $w$ the width of initial Harris current sheet, and $B_{\rm g}$ the guide magnetic field.
    }
    \label{table}
  \end{center}
\end{table}

\begin{figure*}[htbp] 
\centering
    \includegraphics[width=18.0cm]{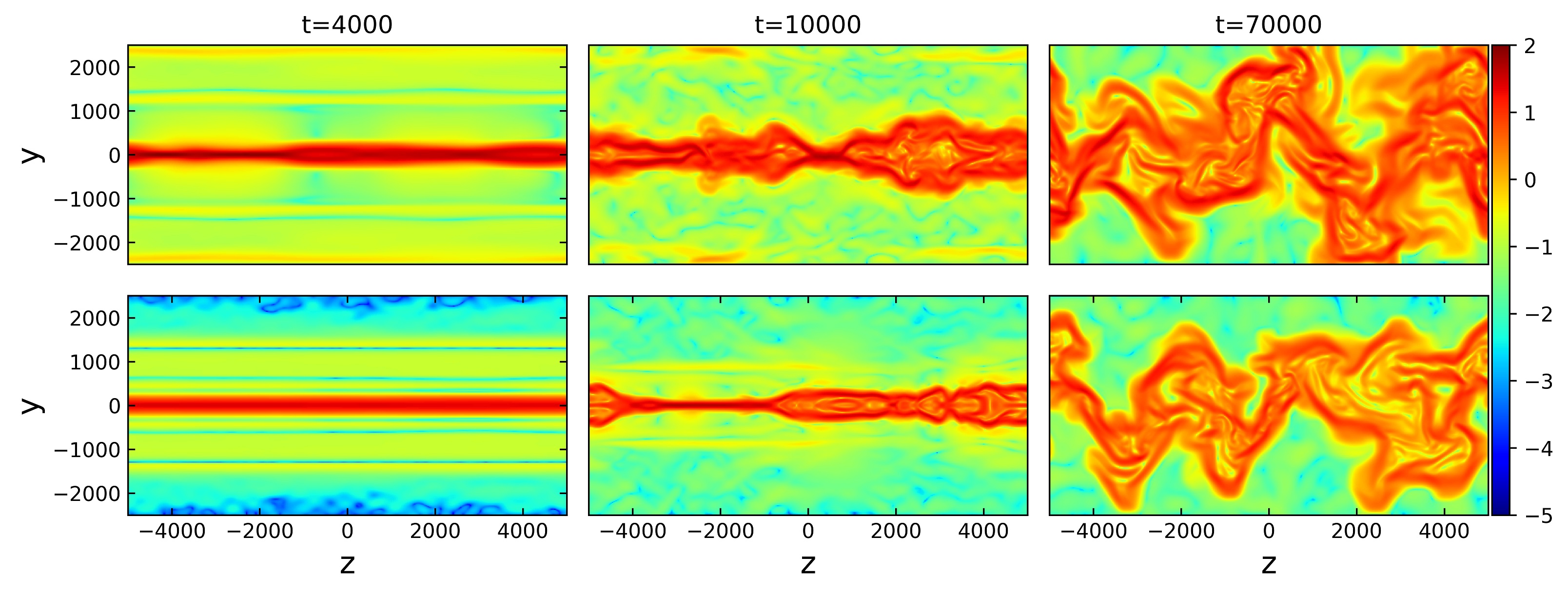}
\caption{The current density from the $Y-Z$ plane at the fixed $X=0$. The color bar represents a logarithm of the current density. The upper and lower panels are based on M1 and M4 listed in Table \ref{table}, respectively. 
} 
    \label{figure-current-width}  
\end{figure*}

\begin{figure}
    \includegraphics[width=8.0cm]{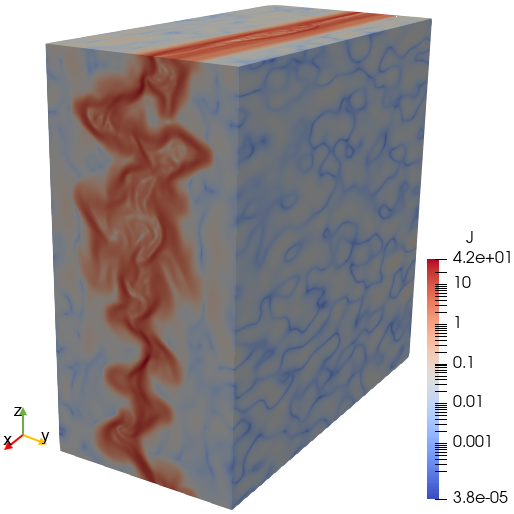}
    \caption{3D visualization of the logarithm of the current density at the $t = 70~000$ in units of the code. The simulation is based on the M2 listed in Table \ref{table}.
    }
    \label{figureJrho} 
\end{figure}

\section{Analysis and results} \label{results}
\subsection{Dynamics of magnetic reconnection} 
We run four models with the initial parameters listed in Table \ref{table} by adding the stochastic noise of velocity fluctuations to launch a magnetic reconnection process. 
We first explore how the current sheet evolves during the spontaneously driven turbulent reconnection process. As an example, Figure \ref{figure-current-width} shows the current density slices on the $Y-Z$ plane at the fixed $X=0$ for the selected three snapshots $t = 4~000, 10~000$ and $70~000$ in units of the code. 
We find that the current density with noise structure was uniformly distributed within a thin current sheet at the early stage of evolution and then begins to form high-density clumps ($t<4~000$). 
When the fluid evolves to $t\sim 10~000$, we see prominent tearing of the current sheet. There are some magnetic islands-like structures similar to 2D magnetic reconnection simulation (e.g., \citealt{Loureiro2007,Mignone2018,Puzzoni2021}). 
This is because the initial current sheet with noise structure provides an instability incentive, which develops various fluid instabilities (e.g., Kelvin-Helmholtz and tearing instabilities) that make the current sheet fragment. 
\cite{Kowal2020} found that tearing instability dominates turbulence generation in the early stages of the evolution, and due to the suppression of tearing instabilities by magnetic field components perpendicular to the current sheet, the dominant mechanism in the later stages shifts to Kelvin-Helmholtz instabilities. Although quantitative analysis of instability is beyond the scope of our paper, we think that similar mechanisms, i.e., tearing and Kelvin-Helmholtz instabilities, should work at the hybrid scales between MHD and plasma scales, which needs to be further confirmed.
As the fluid further evolves, the current sheet begins to thicken and form a significant large-scale magnetic flux rope structure at $t=70~000$. In our simulation, the setup of reflective boundary conditions in the $Y$-direction will cause the magnetic flux to reflect at the boundaries. 
The rebounded magnetic flux enters the reconnection region and intertwines with the magnetic flux in the reconnection region, thus forming a large-scale magnetic flux rope structure. 
This is different from the case of outflow boundary conditions applied in the direction of the perpendicular current sheet plane, in which such large-scale interactions of magnetic fluxes are not allowed (see \citealt{Kowal2017}). 

The structure of the current sheet can also be observed from the 3D perspective in Figure \ref{figureJrho} which shows a 3D visualization of the logarithm of the current density obtained from M2 at $t = 70~000$ in units of the code. It is noticed that the current sheet on the $Y-Z$ plane has obvious tearing, while only thickening on the $X-Y$ plane. The reason is that the magnetic field direction is set along the $X$-axis direction. In addition, due to magnetic tension, the tearing of the current sheet is suppressed on the $X-Y$ plane.

\begin{figure}
\includegraphics[width=8.5cm]{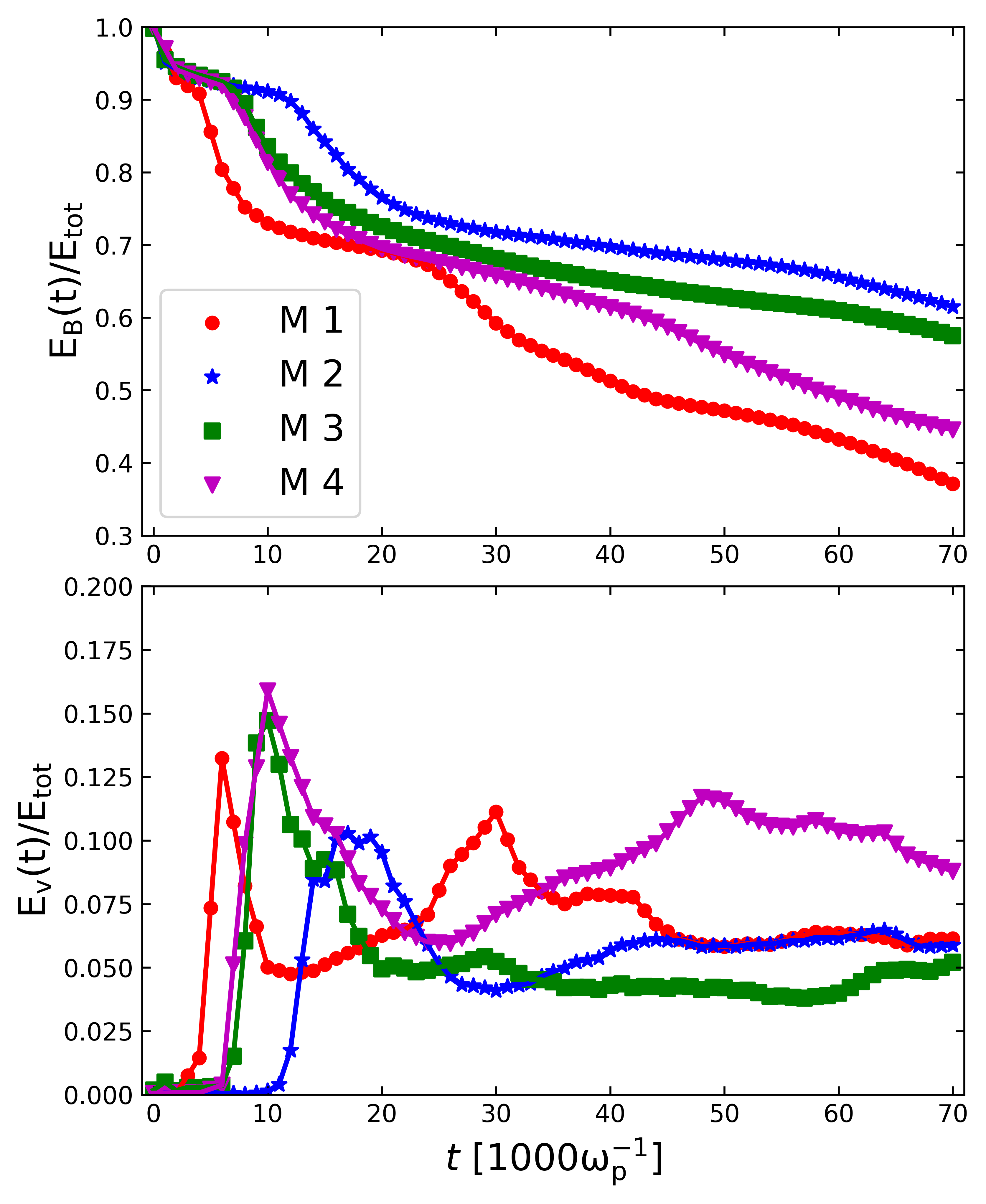}
\caption{Top panel: magnetic energy as a function of time; Bottom panel: kinetic energy as a function of the time. The results of all models are normalized using the initial total energy $E_{\rm tot}(t_0) = E_{\rm v}(t_0)+E_{\rm B}(t_0)$.}
\label{figure-energy} 
\end{figure}
 
Figure \ref{figure-energy} shows the change of magnetic and kinetic energies within the reconnection region over the evolution time. As shown in the top panel of Figure \ref{figure-energy}, the continuously decreased magnetic energy with magnetic reconnection evolution demonstrates the presence of the release of magnetic energy by reconnecting. For model M2, the reduction of magnetic energy always lags behind other models, because its low initial noise added in velocity fluctuations makes its turbulent reconnection slightly slower than other models. Compared with models M3 and M4, the reduction in magnetic energy from model M4 is more significant, especially in the later stage ($t\gtrsim 40~000$). This is because the guiding field makes the magnetic field lines on both sides of the current sheet no longer strictly parallel so that they have a certain angle, which tends to enhance the reconnection efficiency. 
Compared with models M1 and M4, the release of magnetic energy in model M1 is more efficient. Although, this is consistent with conventional thinking: a large $\sigma$ value, i.e., a strong magnetic field, leads to the release of more reconnected magnetic energy. 
However, the more fundamental reason is that instabilities of initial noise excitation make the current sheet thicker (seems to touch near the boundaries) and tear more in the case of a strong magnetic field, as shown in the right upper panel of Figure \ref{figure-current-width}.

The evolution of fluid kinetic energy over time first experiences a short plateau period then increases rapidly to form a sharp peak, and finally gradually tends to a quasi-steady state. 
In the initial stage of the beginning of reconnection, the release efficiency of magnetic energy is low and stochastic noise needs time to excite instabilities that induce the generation of turbulence. 
Subsequently, the magnetic energy is rapidly released, and the kinetic energy is suddenly increased. When the increase and dissipation of fluid kinetic energy reach a balance, the fluid kinetic energy reaches a peak. 
With the development of magnetic reconnection, a part of the dissipated magnetic energy is converted into the kinetic energy of the fluid. As shown in the bottom panel of Figure \ref{figure-energy}, at the M1 peak position, the magnetic energy decreases by about 0.2, while the fluid kinetic energy increases by about 0.13, and the remaining energy is converted into the kinetic energy of test particles. 
Throughout evolution, the average increase in kinetic energy hardly exceeds 0.1, while the continuous release of magnetic energy can reach about 0.6 ($=1-0.4$). 
Furthermore, since there are a few parts of particles in the current sheet at the initial stage of evolution, most of the magnetic energy released by the reconnecting is converted into kinetic energy. 
As the current sheet expands over time, the number of particles in the current sheet will increase, resulting in the direct conversion of magnetic energy into test particle kinetic energy.

Comparing these four models, we find that model M2 has a lag, which is consistent with the evolution of magnetic energy, but the lag phenomenon is more prominent. Model M4 has a higher peak value and a faster increase in kinetic energy than M3. The reason is that it has a significantly faster decrease in magnetic energy. Model M1 presents a more efficient increase of kinetic energy because of its stronger magnetic field.
Both deformations of the current sheet and the formation of magnetic flux ropes can lead to changes in the strength of the outflow from the local reconnection region and interact with the developing turbulence. Turbulence increases when reconnection promotes this interaction. In turn, developing turbulence affects the reconnection process, leading to an increase in reconnection rates. With the evolution of time, the region around the current sheet is also successfully turbulent, providing an opportunity for the particles escaping the current sheet or the local particles to interact with turbulence.

\begin{figure*}
\centering
\includegraphics[width=18.0cm]{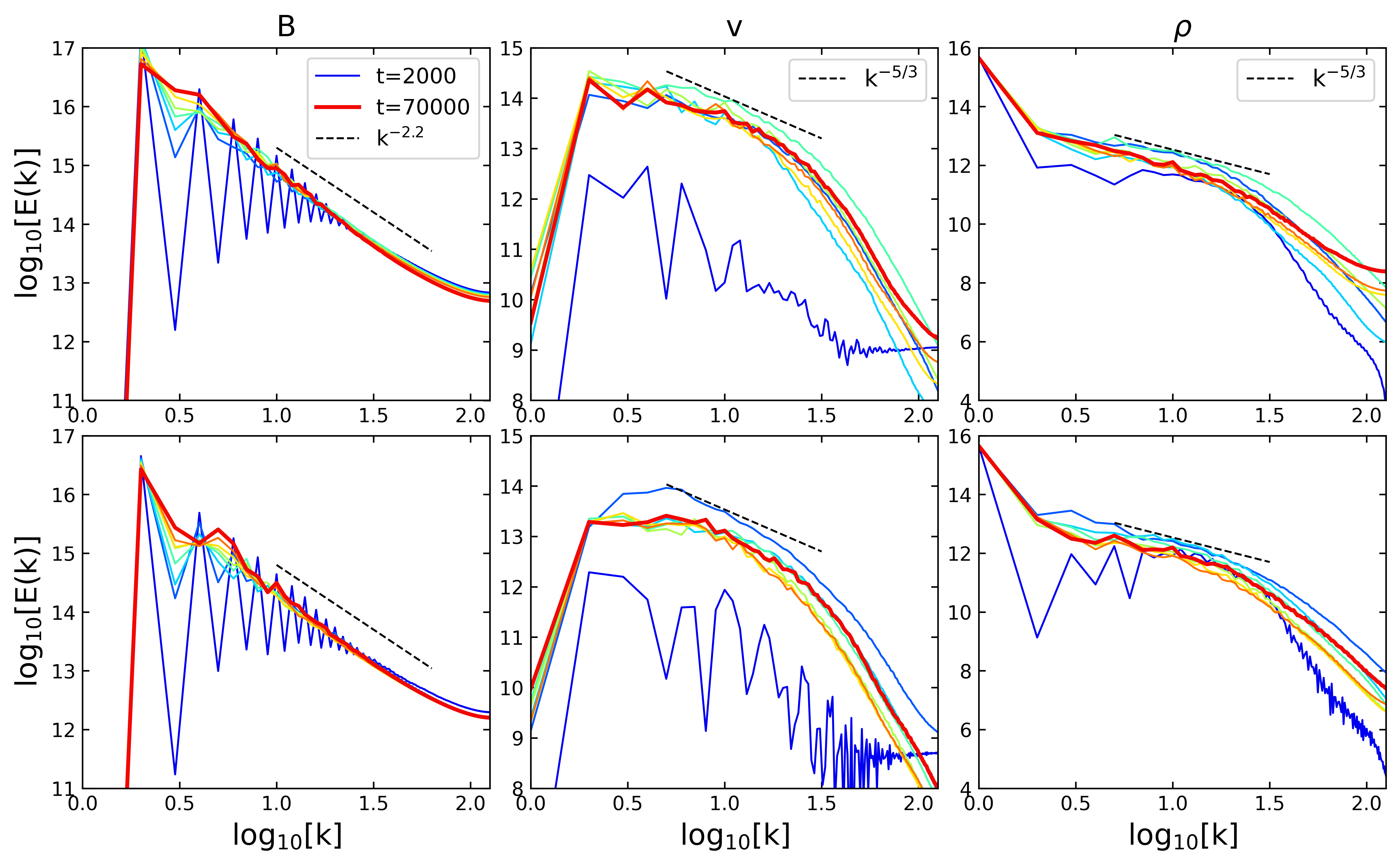}
\caption{The power spectra of magnetic fields (left column), velocities (middle column), and densities (right column) at the different evolution times. The top and bottom rows are based on M1 and M3 listed in Table \ref{table}, respectively. The red thick curves in each panel show the spectra at the final snapshot of $t = 70~000$ in units of the code. 
} 
\label{figure-power-spectra} 
\end{figure*}

\subsection{Properties of reconnection-driven turbulence}
We show the power spectra at the different evolution snapshots in Figure \ref{figure-power-spectra}, including the magnetic fields (left column), velocities (middle), and densities (right) for the model M1 (upper panels) and M3 (lower). These snapshots correspond to $t = 10~000$ to $70~000$ at intervals of $10~000$, where for comparison we also include the power spectrum at the initial stage of evolution of $t = 2~000$. Since turbulence just begins to develop, the power spectrum fluctuates greatly for magnetic fields, velocities, and densities. With the evolution of time, the reconnection rate gradually increases, and the fully evolved power spectrum of MHD turbulence gradually converges. In the case of the velocity and density, they all well present the scaling of $k^{-5/3}$, which is the same as the \cite{Kolmogorov1941} spectrum. 
This demonstrates that reconnection-driven turbulence also provides a scaling of $-5/3$ similar to that of Kolmogorov and GS95. However, the magnetic field presents the scaling of $k^{-2.2}$ as shown in the left panels, which is steeper than the Kolmogorov spectrum. This numerical experimental result implies that magnetic and velocity vector fields do not couple well. In our work, we also calculate the power spectra from models M2 and M4 (not shown) and find no significant differences between them.

\begin{figure} 
\includegraphics[width=8.5cm]{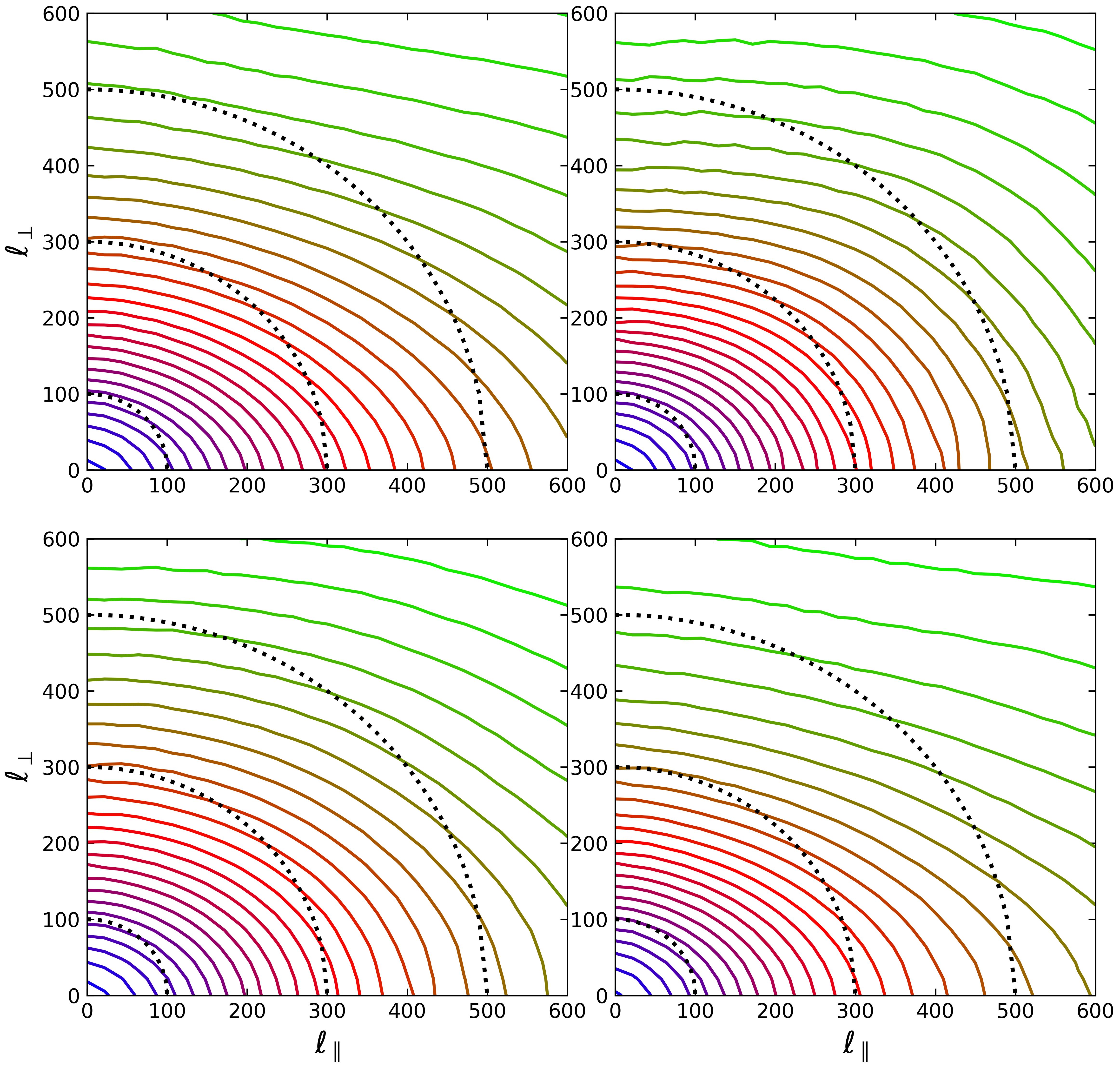}
\caption{The anisotropy of the velocities with respect to the local magnetic fields at the finial snapshot $t = 70~000$ in units of the code. $l_{\perp}$ and $l_\parallel$ represent the perpendicular and parallel scales of the eddies along the local magnetic field directions, respectively. Dotted lines indicate isotropy.
}
\label{figure-sf-map}
\end{figure}

\begin{figure}  
\includegraphics[width=8.5cm]{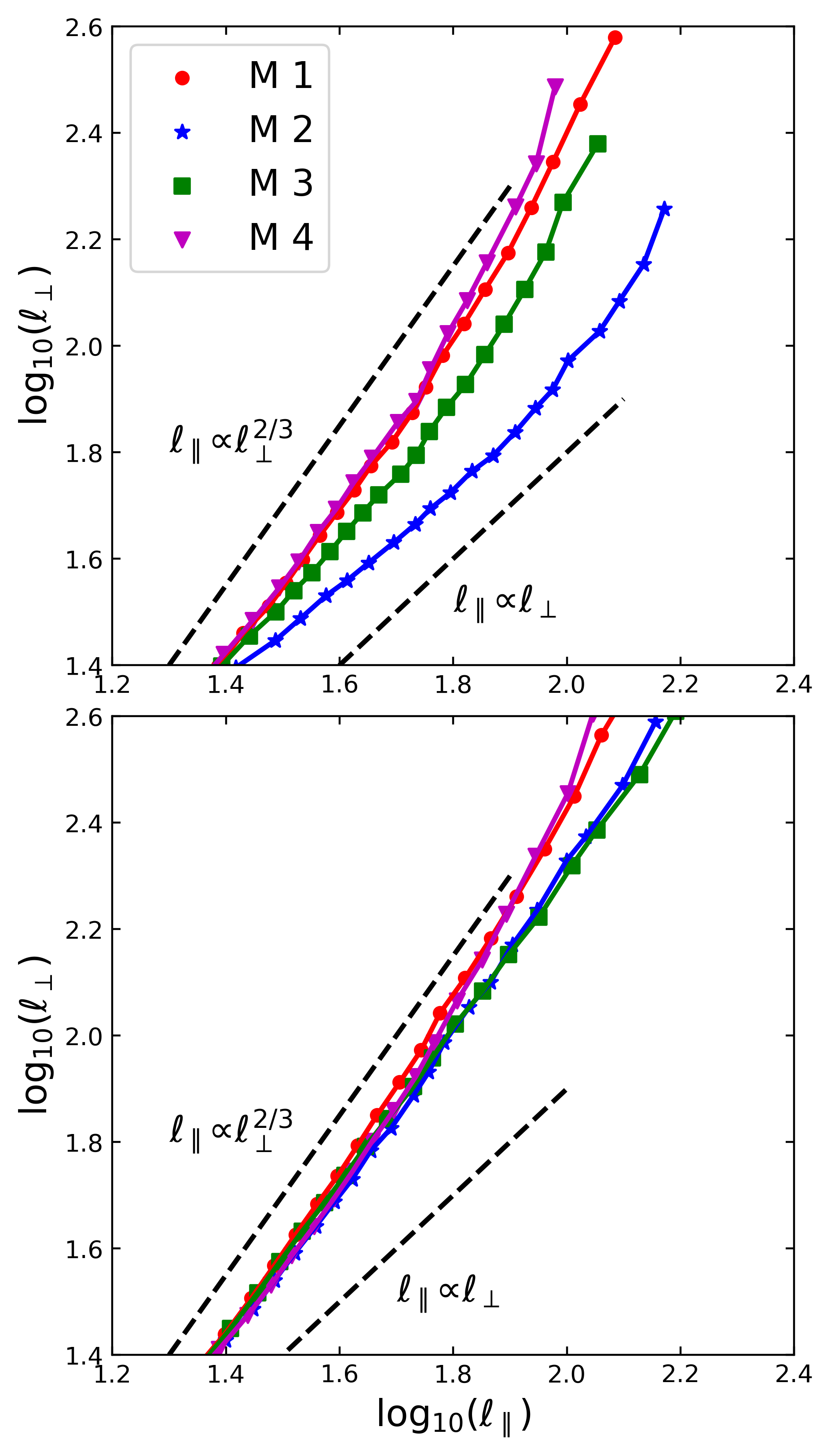}
\caption{Anisotropy scalings of velocities at $t = 30~000$ (upper panel) and $70~000$ (lower panel). $l_{\perp}$ and $l_\parallel$ represent the perpendicular and parallel scales of the eddies along the local magnetic field directions, respectively.
} 
\label{figure-sf-scale} 
\end{figure}

Given the consistency of the velocity spectra with the generally accepted GS95 model, we analyze the anisotropic property of velocity in depth. 
We plot in Figure \ref{figure-sf-map} the contour maps of the second-order structure function of velocity in a local reference frame for four models explored at the final snapshot ($t=70~000$). We can see that in all of these four models, it presents a well-anisotropic property. The semi-major axis is along the $\ell _{\parallel}$ direction, which is parallel to the local mean magnetic field. In addition, the models M1 and M4 show slightly stronger anisotropy than M2 and M3. This may be because the two formers correspond to a strong mean magnetic field and an additional guiding field, respectively, both of which have stronger magnetic field effects.

To quantify the anisotropy of reconnection-driven turbulence, we also show the anisotropy scaling of velocity in Figure \ref{figure-sf-scale}, arising from the early ($t = 30~000$) and late ($t = 70~000$) snapshots. For the early snapshot (upper panel), the models M1 and M4 are very similar, both of which will satisfy the GS95 scale-dependent anisotropy of $\ell_\parallel \propto \ell_{\perp}^{2/3}$ in the small-scale region, while the model M2 tends to $\ell_\parallel \propto \ell_{\perp}$ and M3 is in between. This indicates that a strong mean magnetic field and guiding field will significantly promote the development of reconnection-driven turbulence. As for the late snapshot (lower panel), model M1 is similar to M4, while model M2 to M3. In the large-scale region, the anisotropic scaling of the models M2 and M3 are slightly smaller than that of M1 and M4, which is consistent with the results shown in Figure \ref{figure-sf-map}. We found that when self-generated turbulence reaches a statistically steady state (terminated at $t = 70~000$), four models all are presenting the GS95 scale-dependent anisotropy in the inertial range of turbulence cascade. 

\subsection{Particle acceleration} 

\begin{figure}
\includegraphics[width=8.5cm]{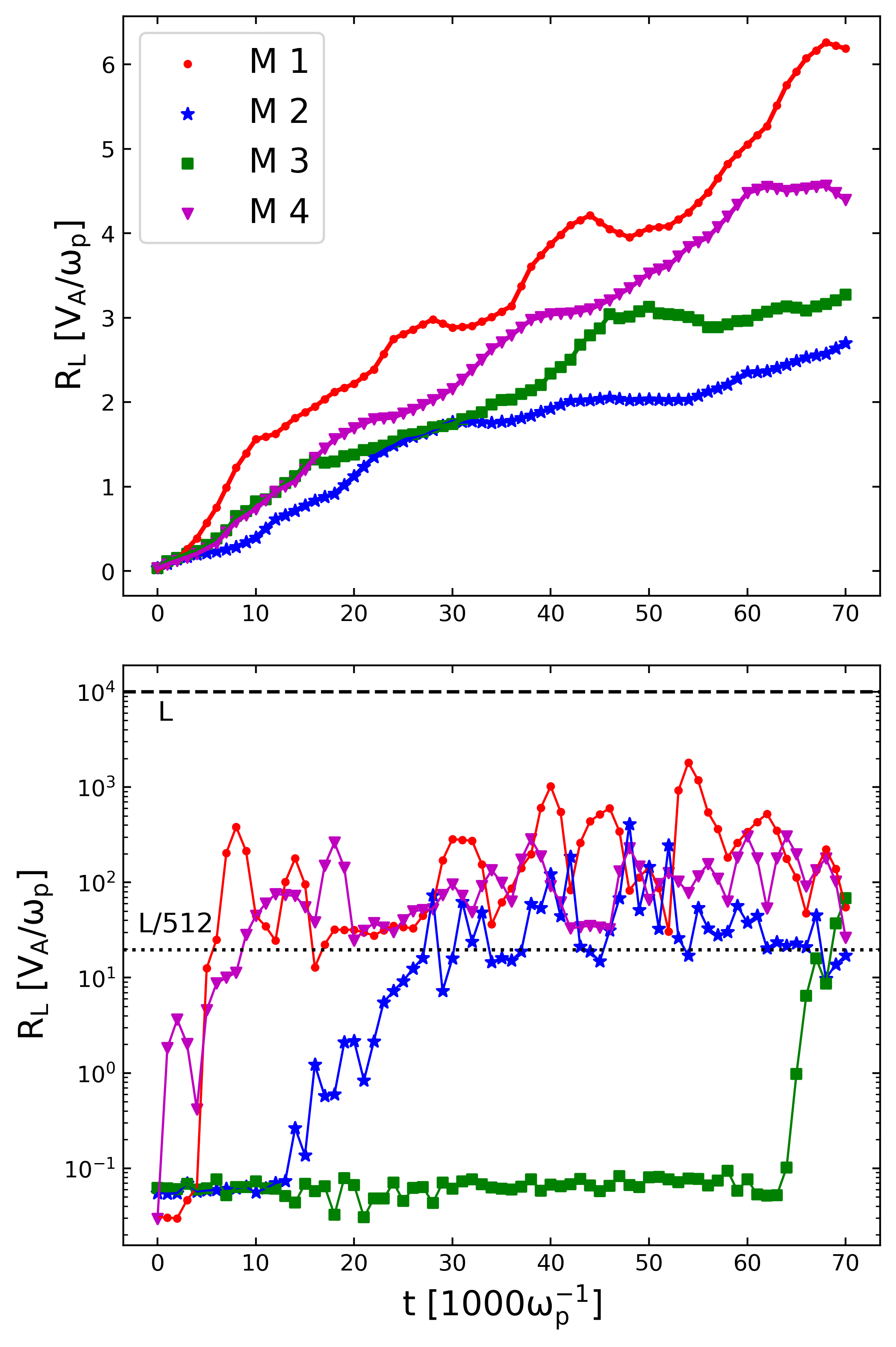}
\caption{The gyroradius of particles as a function of the time: all particles (top panel) and four particles selected (bottom panel, based on model M2). The red and purple curves on the bottom panel indicate the evolution of the particles inside the initial current sheet, while the blue and green curves denote around and away from the current sheet, respectively. The horizontal dotted line plotted on the bottom panel represents the grid size of $ \delta L \approx 20$, and the horizontal dashed line represents the boundary of the box.
}
\label{figure-rl} 
\end{figure}

In this section, we explore the evolution of CRs over time in reconnection-driven turbulence, including the gyroradius, momentum diffusion coefficient, and spectral energy distribution. 
The gyroradius of particles evolving with time is shown in Figure \ref{figure-rl} containing the average gyroradius of all particles (upper panel) for four models and four accelerated particles randomly selected (lower panel) in model M2. We can see that for all four models the gyroradius increase gradually with time, and the difference is that the growth rate of models M1 and M4 is higher than that of the other two models. 
This should be due to the effect that both a high-$\sigma$ parameter for M1 and a non-zero guide field for M4 can improve the reconnection rate to accelerate particles more efficiently.
Moreover, the growth rate of the gyroradius of model M1 (high-$\sigma$ setup) is significantly higher than that of M4. 
This implies that the guide field is not the only factor that improves the reconnection rate. For the phenomenon of the lowest growth rate of model M2, it can attribute to the addition of low initial noise, which makes turbulent reconnection slightly slower. 

The lower panel of Figure \ref{figure-rl} plots the time evolution of the gyroradius of selected four particles: two of them inside the initial current sheet (red and purple curves), one around (blue curve), and away from (green curve) the current sheet. For particles within the current sheet, sufficient energy is obtained in the early stages, and its gyroradius is increased by about 3-4 orders of magnitude, while the acceleration time is delayed for the particle around and away from the current sheet. It is not difficult to understand that as magnetic reconnection develops, the current sheet becomes wider and wider, and more particles are involved in acceleration.

\begin{figure}
\includegraphics[width=8.5cm]{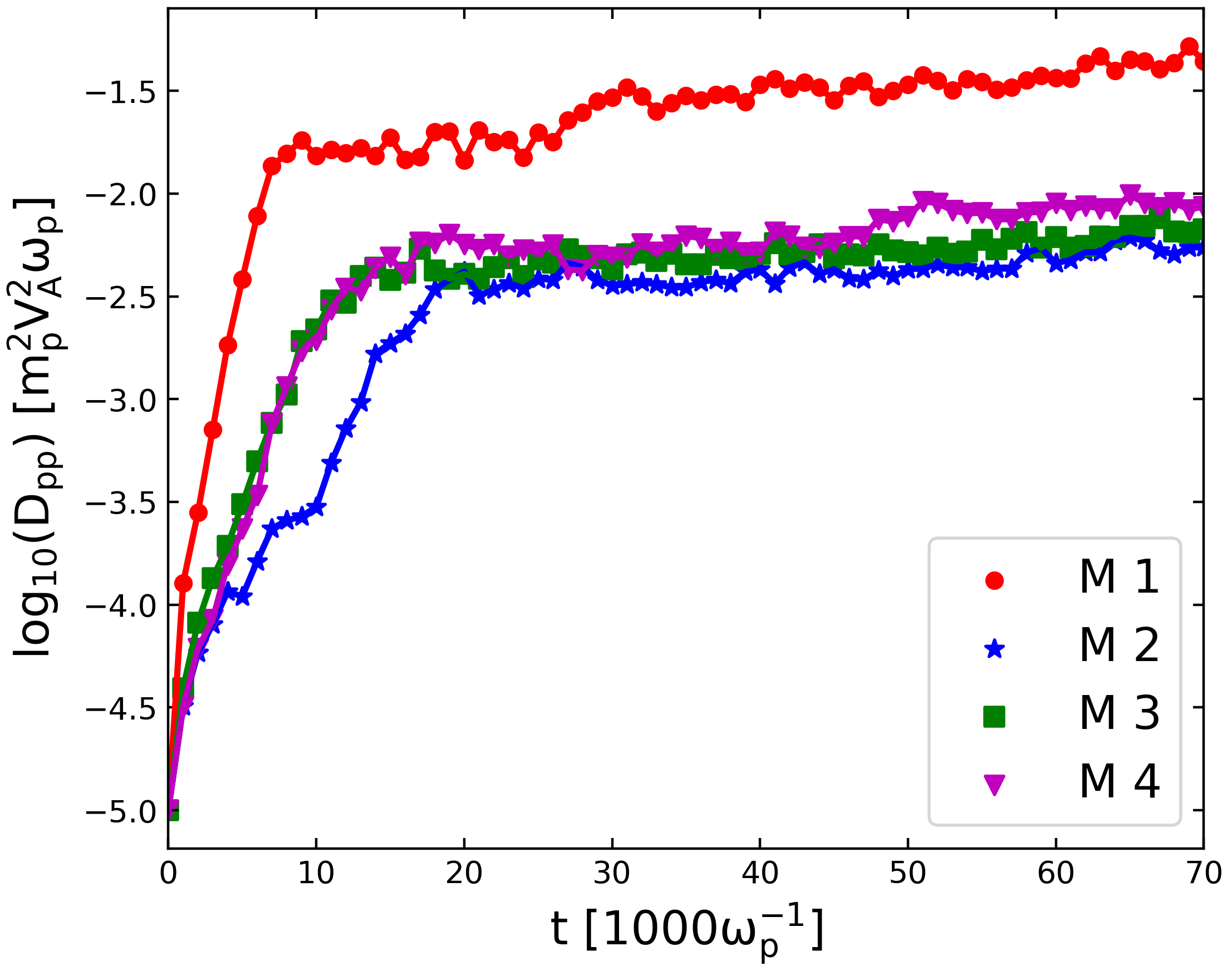}
\caption{Variation of particle momentum diffusion coefficient over time. 
}
\label{figure-dpp}
\end{figure}

Figure \ref{figure-dpp} shows the evolution of the momentum diffusion coefficient over time for the four models. From this figure, we can see a rapid growth of $D_{\rm pp}$ in the early stage, followed by a quasi-steady state as a plateau (at $t\sim 10~000$ for the model M1, while other models are a little bit later $t\sim 20~000$). 
In general, the plateau characteristics in $D_{\rm pp}$ are usually regarded as a typical second-order Fermi process (\citealt{Pezzi2022,Gao2023}). However, the current turbulent reconnection situation is more complicated because the acceleration time of individual particles is inconsistent. The calculation we provide here is statistical averaging of all particles, just to observe the diffusion behavior of particles in the reconnection region. In particular, the momentum diffusion in M1 exhibits differences from other models. This suggests that the main factor influencing momentum diffusion is the strength of the magnetic field, and the second factor may be linked to small initial velocity perturbations, the width of the current sheet, and the additional guide field.

\begin{figure}
\includegraphics[width=8.5cm]{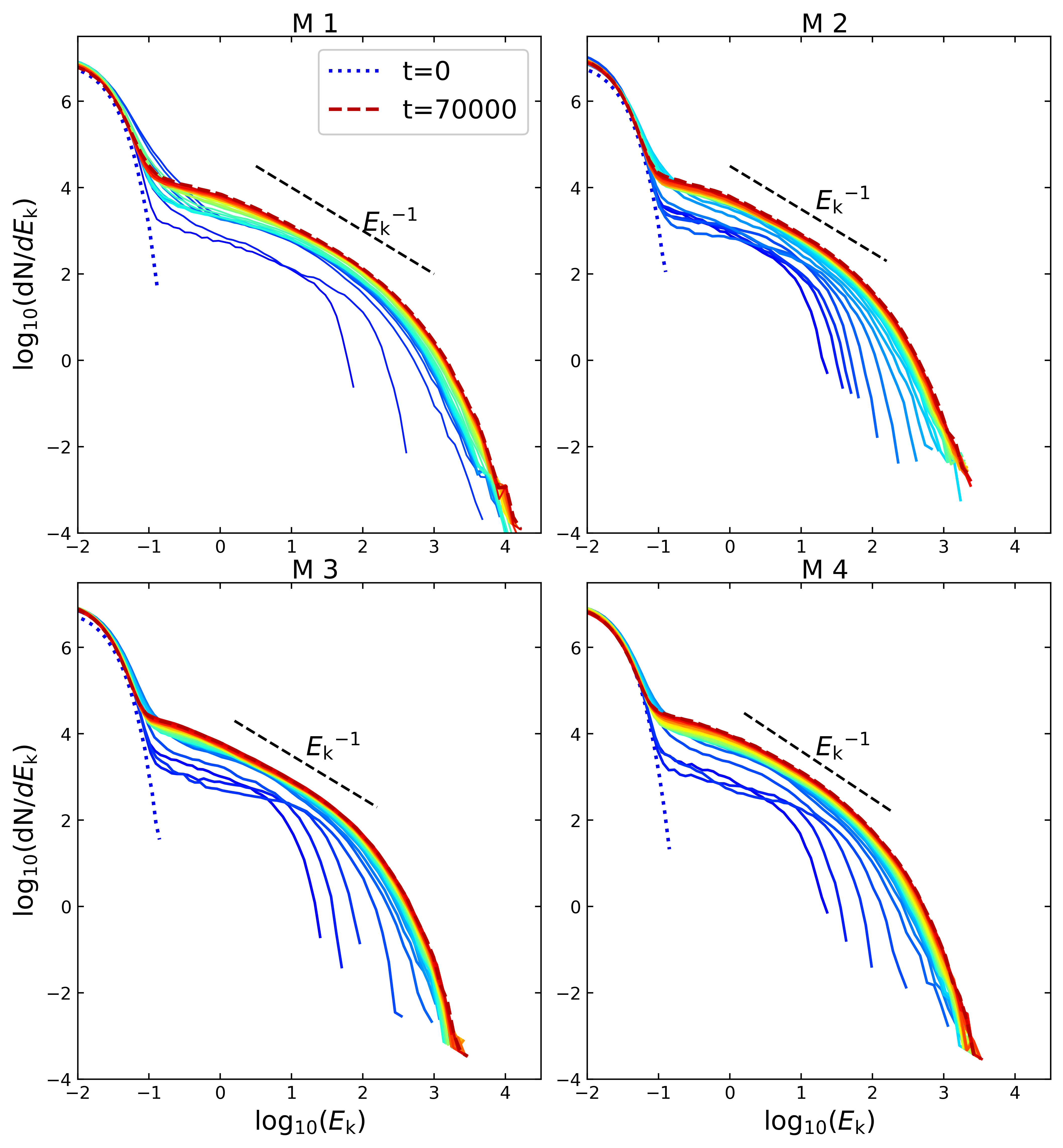}
\caption{Spectral energy distributions of all particles at the different evolution times for models M1-M4. The particle distributions at the initial and final snapshots are marked with dotted and dashed lines, respectively.
} 
\label{figure-energy-spectrum} 
\end{figure}

Furthermore, Figure \ref{figure-energy-spectrum} shows spectral energy distributions of one million particles injected for the four models at 36 snapshots from $t = 0$ to $70~000$ with an interval of $2~000$. 
From this figure, we can find the following main characteristics. First of all, as time evolves, the particle spectra from these models are broadened and moved to higher energy from the initial thermal distributions to the non-thermal ones. Their distributions at the final snapshot almost show a hard spectrum with the power-law relationship of $dN/{{\rm d}E_{\rm k}} \propto E_{\rm k}^{-1}$, which is consistent with the results of PIC simulations \citep{Guo2014, Guo2015}. Secondly, the spectral energy distributions enter the high-energy cut-off, because only a few fractions of the particles can be accelerated to such high energy. In addition, an important difference between the four models is that the maximum acceleration energy of model M1 (upper-left panel) reaches above $10^4$, while the other three models are around $10^{3.5}$. This indicates that the strong mean magnetic field promotes the reconnection acceleration of particles, allowing the accelerated particles to reach a higher energy upper limit.

\begin{figure*}
    \centering
    \includegraphics[width=18.0cm]{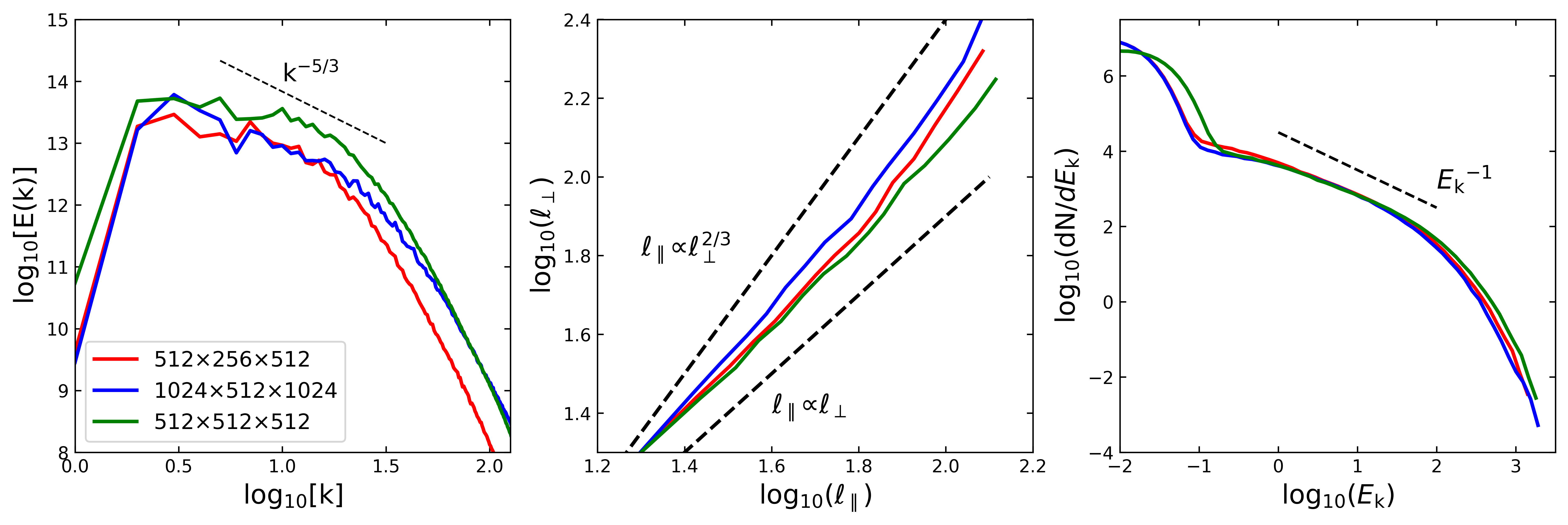}
    \caption{The power spectra of velocities (left panel), anisotropy scalings of velocities (middle panel), and particle energy spectra (right panel) for the model M3 with different resolutions obtained at $t = 30~000$.}
    \label{figure-numerical-resolution}
\end{figure*}

\subsection{Comparison of different numerical resolutions} \label{subsection 5.3}
To verify the reliability of the numerical results, we run two high-resolution models (one is $1024\times 512 \times 1024$, and another is $512^3$) and compare the numerical results, including the power spectra (left panel) and anisotropy scalings of velocity (middle), as well as particles energy spectra (right) in Figure \ref{figure-numerical-resolution}. 
Except for the resolution, other parameters of these models are kept consistent with model M3. Because of the limited computing resources, we only simulate it up to $t = 30~000$, at which we see that the power spectra of velocities all show a Kolmogorov-like spectra $E(k) \propto k^{-5/3}$, and the spectral energy distributions of particles present as the hard spectrum with an exponent of $-1$. 
As shown in the middle panel, the phenomena that the anisotropic scale does not satisfy $\ell_{\parallel} \propto l_{\perp}^{2/3}$ but more conforms to $\ell_{\parallel} \propto l_{\perp}$ is due to the insufficient run time. 
When the evolution time is sufficiently long, it should show more significant anisotropy. Nevertheless, the numerical resolution has little effect on our results, and the simulation settings in this paper are sufficient for our moderate aim. 

\section{Discussion} \label{Discussion}
At present, many studies are using MHD plus test particles or PIC methods. The former mostly focused on large MHD scales, which ignores the kinetic effects associated with the microscale (e.g., \citealt{Gordovskyy2010a,Gordovskyy2010b,Kowal2011}), while the latter focused on small plasma scales, which in most cases are several orders of magnitude smaller than the overall size of typical astrophysical systems (e.g., \citealt{Guo2015,Guo2021,Mignone2020}). In this work, we carried out the numerical simulations of self-driven turbulent magnetic reconnection, and explored the properties of reconnection-driven turbulence and reconnection acceleration on the transition scale between the large MHD scale and small kinetic plasma one, by using the MHD-PIC method and injecting $10^6$ particles into our simulation box.

We applied the initial Harris current configuration (\citealt{Puzzoni2021, Mignone2020}) and let a noise added in velocity fluctuations affect the current sheet with an initial magnetic field discontinuity (e.g., \citealt{Beresnyak2017,Kowal2017}). Differently, we did not constrain particles to a frozen 3D MHD snapshots like \cite{Liu2009}, \cite{Kowal2011}, \cite{Gordovskyy2010a} and \cite{Ripperda2017a}, but make particles and fluid coevolution (e.g., \citealt{Gordovskyy2010b, Ripperda2017b,Mignone2020}). 
Moreover, we use reflective boundary conditions in the $Y$ direction, while periodic boundary conditions in the $X$ and $Z$ directions, which is different with \cite{Kowal2017} (outflow boundary in the direction perpendicular to the current sheet) and \cite{Beresnyak2017} (periodic boundary in the direction perpendicular to the current sheet). 
This different set of reflective boundary conditions has some important influences. Most notably, the magnetic flux is reflected into the current sheet at the boundary, thus forming a significant large-scale magnetic flux rope structure, as shown in Figure \ref{figure-current-width}. 
The periodic boundary conditions in all three directions that \cite{Beresnyak2017} used make the horizontal large-scale interaction between the two current sheets exist and the large-scale motions can be generated in the vertical direction. 
While such large-scale interactions are not allowed in the case of \cite{Kowal2017}.

As we mentioned in Section \ref{results}, the properties of reconnection-driven turbulence are in perfect agreement with \cite{Kolmogorov1941} and GS95, such as the scaling of $k^{-5/3}$ for velocity and density. The contour map of the second-order structure function of velocity presents anisotropic properties, quantitatively satisfying the relationship of $\ell_\parallel \propto \ell_{\perp}^{2/3}$ in the inertial range. It should be noted that the variable parameters used in our four models are all related to the magnetic field, except for the amplitude of initial velocity perturbation $V_{\rm eps}$. For the acceleration behavior of particles in the reconnection region, particles are accelerated all the time in evolution. Although the momentum diffusion coefficient except for the early stages of evolution shows a characteristic of the second-order Fermi acceleration with time \citep{Pezzi2022}, we avoid stating what kind of acceleration mechanism it is, which will be left to future work.

When studying the properties of the turbulence generated in the reconnection region, we found that the power spectrum of the magnetic field exhibits a scaling of $k^{-2.2}$, which is steeper than the GS95 scaling of $k^{-5/3}$, indicating that the magnetic energy is mostly concentrated in the small-scale region in the reconnection-driven turbulence. It is different from the traditional cascade of energy transfer from large scale to small scale by an external force, which may be associated with the inverse cascade process. In the current work, we did not explore the properties of the decomposed modes of reconnection-driven turbulence, which needs to be further investigated. At the same time, we did not consider the Hall effect and particle feedback. For the former, although the effect of Hall-MHD on turbulent reconnection at a scale larger than the ion skin depth can be negligible (\citealt{Beresnyak2018}), it may be important for the reconnection at a sufficiently small scale (\citealt{Lazarian2020}). For the latter, it may affect the particle acceleration efficiency and thus change the spectral energy distribution of accelerated particles.

\section{Summary} \label{Summary}
By using the MHD-PIC method, we performed 3D numerical simulations of magnetic reconnection-driven turbulence and turbulent reconnection acceleration of particles. In the framework of the coevolution of fluid and particles, we focus on exploring the dynamics of magnetic reconnection processes, analyzing the properties of self-driven turbulence, and understanding the acceleration behavior of particles. The main results are briefly summarized as follows.

\begin{enumerate}
    \item We find that magnetic reconnection can spontaneously generate turbulence in the range of the transition scale between the large MHD scale and the small kinetic plasma one. The deeper reason is instability leading to changes in the structure of the reconnecting layer and the generation of turbulence.
   
    \item By the time the self-driven turbulence reaches a statistically steady state, the magnetic energy is almost released by 50\%, while the kinetic energy of the fluid increases by no more than 15\%.
    
    \item In the case of the velocity and density fluctuations, the reconnection-driven turbulence is similar to the turbulence driven by an external force, that is, they present the scaling index of $-5/3$ in the inertial range and the anisotropy relationship of $\ell_{\parallel} \propto l_{\perp}^{2/3}$, in agreement with the GS95 theory.

    \item In the case of magnetic field fluctuations, the reconnection-driven magnetic turbulence presents a power spectral index of $-2.2$, which is steeper than the turbulence driven by an external force. This implies that the most of magnetic energy cascade in the self-driven turbulence occurs in the small-scale region.
    
    \item The strong magnetic field tends to enhance the turbulent reconnection efficiency to accelerate particles more efficiently. The non-zero guiding field can significantly improve turbulent magnetic reconnection processes.
    
    \item The self-driven turbulence can enhance the reconnection efficiency, resulting in a wider current sheet and a significant magnetic flux rope structure. This effectively improves the acceleration of the particles, providing a hard particle spectral distribution of $N\propto E^{-1}$.

    
  \end{enumerate}

\begin{acknowledgments}
We thank the anonymous referee for valuable comments that significantly improved the quality of the paper. We thank the support from the National Natural Science Foundation of China (grant No. 11973035), the Hunan Province Innovation Platform and Talent Plan–HuXiang Youth Talent Project (No. 2020RC3045), and the Hunan Natural Science Foundation for Distinguished Young Scholars (No. 2023JJ10039).
\end{acknowledgments}

\vspace{5mm}

          
\bibliography{sample631}{}
\bibliographystyle{aasjournal}

\end{document}